\documentclass[a4paper, 11pt]{article}

\usepackage{amsmath}
\usepackage{amssymb}
\usepackage{amsthm}
\usepackage{amsfonts}
\usepackage{amstext}
\usepackage{amsopn}
\usepackage{amsxtra}
\usepackage[latin1]{inputenc}
\usepackage{dsfont}
\usepackage{bm}

\newtheorem{thm}{Theorem}
\newtheorem{lemma}{Lemma}

\theoremstyle{definition}

\newtheorem{remark}{Remark}

\newcommand\R{{\ensuremath {\mathbb R} }}
\newcommand\C{{\ensuremath {\mathbb C} }}

\newcommand\Z{{\ensuremath {\mathbb Z} }}

\newcommand\1{{\ensuremath {\mathds 1} }}

\renewcommand\phi{\varphi}
\newcommand{\gH}{\mathfrak{H}}
\newcommand{\gS}{\mathfrak{S}}

\newcommand{\wto}{\rightharpoonup}
\newcommand{\cP}{\mathcal{P}}

\newcommand\ii{{\ensuremath {\infty}}}
\newcommand\pscal[1]{{\ensuremath{\langle #1 \rangle}}}
\newcommand{\norm}[1]{ \left| \! \left| #1 \right| \! \right| }
\newcommand\ket[1]{{\ensuremath{\left|#1\right\rangle}}}
\newcommand\bra[1]{{\ensuremath{\left\langle#1\right|}}}
\def\tr{\mathop{\rm tr}\nolimits} 
\newcommand{\cE}{\mathcal{E}}

\newcommand{\cK}{\mathcal{K}}
\newcommand{\cB}{\mathcal{B}}
\newcommand{\cC}{\mathcal{C}}
\newcommand{\cS}{\mathcal{S}}

\newcommand{\exc}{X}

\newcommand \dps{\displaystyle }

\begin{document}
\title{TITLE}
\author{Marco GHIMENTI \& Mathieu LEWIN}

\begin{center}
\Large \textbf{Properties of the periodic Hartree-Fock minimizer}
\end{center}

\begin{center}
 \large Marco GHIMENTI$^a$ \& Mathieu LEWIN$^b$
\end{center}

\medskip

\begin{center}
\small

 $^a$Dipartimento di  Matematica Applicata ``Ulisse Dini", Università di Pisa,
 Via Buonarroti, 1/c - 56127 - Pisa,  ITALY. (Partially supported by INDAM)

\texttt{ghimenti@mail.dm.unipi.it}

\medskip

$^b$CNRS \& Department of Mathematics UMR8088, University of Cergy-Pontoise, 2 avenue Adolphe Chauvin, 95302 Cergy-Pontoise Cedex, FRANCE.

\texttt{Mathieu.Lewin@math.cnrs.fr}
\end{center}

\begin{center}
 \it March 22, 2008
\end{center}

\smallskip

\begin{abstract}
We study the periodic Hartree-Fock model used for the description of electrons in a crystal. The existence of a minimizer was previously shown by Catto, Le Bris and Lions (\textit{Ann. Inst. H. Poincar\'e Anal. Non Lin\'eaire} \textbf{18} (2001), no.~6, 687--760). We prove in this paper that any minimizer is necessarily a projector and that it solves a certain nonlinear equation, similarly to the atomic case. In particular we show that the Fermi level is either empty or totally filled.
\end{abstract}

\section{Introduction}
The Hartree-Fock model is widely used to describe quantum electrons in usual matter. The purpose of this paper is to provide interesting properties of the Hartree-Fock ground state in the periodic case, that is to say in a condensed matter setting.

In Hartree-Fock theory \cite{LS,BLS}, the state of the electrons is represented by a so-called \emph{density matrix} $\gamma$. This is a self-adjoint operator
$0\leq\gamma\leq1$ acting on the physical space $L^2(\R^3)$. When $\gamma$ has a
finite trace, it models a finite number of electrons. An important example is given by a so-called \emph{Hartree-Fock state}
\begin{equation}
\gamma=\sum_{n=1}^N\ket{\phi_n}\bra{\phi_n}
\label{form_proj_density_matrix}
\end{equation}
which is an orthogonal projector of rank and trace $N$, $(\phi_n)_{n=1}^N$ being an orthonormal basis of the range of $\gamma$. This example models $N$ uncorrelated electrons, i.e. it corresponds to an $N$-body wavefunction of the form $\Psi=\phi_1\wedge\cdots\wedge\phi_N$ in the antisymmetric product $\bigwedge_1^NL^2(\R^3)$. The only correlation present in such a wavefunction is that of the Pauli principle, i.e. the fact that $\Psi$ is antisymmetric.

The Hartree-Fock energy is a nonlinear functional defined for any density matrix $\gamma$ and it has been largely studied in the literature \cite{LS,Lions,BLS,Lieb,BLLS}. An important result due to Lieb \cite{Lieb} is that the minimization of this functional over all possible density matrices having trace $N$, gives the same minimum as when it is restricted to projectors of the form \eqref{form_proj_density_matrix}. Additionally, any minimizer is automatically a projector, $\gamma=\sum_{n=1}^N\ket{\phi_n}\bra{\phi_n}$, and $\phi_1,...,\phi_N$ are the first eigenfunctions of a self-adjoint operator $H_\gamma$ depending on $\gamma$
\begin{equation}
 H_\gamma\phi_n=\lambda_n\phi_n,
\label{form_eq_orbitales_intro}
\end{equation}
$\sigma(H_\gamma)=\{\lambda_1\leq\cdots\leq\lambda_N\leq\cdots\}\cup[0,\ii)$.
Notice that projectors of rank $N$ form the extremal points of the convex set of all density matrices of trace $N$. Although it may seem natural that a minimizer is always an extremal point of the variational convex set, this is indeed a non trivial result as the Hartree-Fock energy is itself \emph{not} concave.
All these properties are important for the device of efficient numerical methods \cite{CLB,book}.

It was shown in \cite{BLLS} that necessarily $\lambda_N<\lambda_{N+1}$ which means that we can write the nonlinear equation satisfied by a minimizer in the form
$$\gamma=\chi_{(-\ii,\lambda_N]}(H_\gamma).$$
Throughout this article, we denote by $\chi_I$ the
characteristic function of the set $I\subset\R$ and by $\chi_{I}(A)$ the
spectral projector on $I$ of the self-adjoint operator $A$.

\medskip

This paper is devoted to the study of the periodic case, modeling an infinite quantum crystal. Our main goal is to prove that any Hartree-Fock minimizer satisfies properties similar to the molecular case described before. For the sake of simplicity, we assume that the crystal is simply the lattice $\Z^3$ and that there is only one nucleus of charge $Z$ at each site of $\Z^3$. It is straightforward to generalize our result to any other periodic system.

In the periodic case studied in the present paper, a state of the system is also described by a density matrix $\gamma$. But this is no more a trace-class operator (it describes infinitely many electrons in the crystal) and it commutes with the translations of the lattice. As before, the orthogonal projectors, the extremal points of the set of all density matrices, will play a special role.

In the periodic setting, there is also a periodic Hartree-Fock-like energy functional depending on $\gamma$, see Formula \eqref{def_energy} below. It was proved by Catto, Le Bris and Lions \cite{CLL} that this energy admits a minimizer, but very few properties of this minimizer are known at present. We prove in this paper a result similar to \cite{Lieb, BLLS}. Namely we show that any minimizer $\gamma$ of the periodic Hartree-Fock energy is indeed \emph{always a projector} and that it solves an equation of the form
\begin{equation}
 \gamma=\chi_{(-\ii,\mu)}(H_\gamma)+\epsilon\chi_{\{\mu\}}(H_\gamma)
\label{eq:SCF_intro}
\end{equation}
with $\epsilon\in\{0,1\}$. The spectrum of $H_\gamma$ is composed of bands and $\mu$ may be an eigenvalue (of infinite multiplicity, due to the invariance by translations of the lattice). Hence the nonlinear equation \eqref{eq:SCF_intro} solved by a minimizer cannot \emph{a priori} be written in a simpler form like \eqref{form_eq_orbitales_intro}.

In \cite{CDL}, a result similar to \eqref{eq:SCF_intro} was recently proved for a simpler model where the only non-convex term, called the \emph{exchange term}, is neglected. In this case the proof is much simpler than in the general case treated in the present article. It is even possible to show that the spectrum of the self-adjoint operator $H_\gamma$ is purely absolutely continuous, hence $\mu$ cannot be an eigenvalue and one can take $\epsilon=0$ in \eqref{eq:SCF_intro}.

\section{Model and main result}
For the sake of simplicity, we assume that the nuclei are placed on the lattice $\Z^3$ and that they all have a charge $+Z$. We also discard the spin of the electrons. Our result can of course easily be generalized to any periodic system of spin-$1/2$ particles.

\subsection{Notation}
Let $\gH$ be a Hilbert space. We denote by $\cB(\gH)$ the space of bounded operators and by $\cS(\gH)$ the set of (possibly unbounded) self-adjoint operators acting on $\gH$. In the whole paper, we denote by $\gS_p(\gH)$
the Schatten class of operators $Q$ acting on the Hilbert space $\gH$ and having a
finite $p$ trace, i.e. such that $\tr_{\gH}(|Q|^p)<\ii$. Note that $\gS_1(\gH)$ is
the space of trace-class operators on $\gH$, and that $\gS_2(\gH)$ is the space of
Hilbert-Schmidt operators on $\gH$. We note by $L^2(M)$ the space of square-integrable complex-valued functions on the Borel set $M$.
Let us also introduce $\Gamma =[-1/2,1/2)^3$ the unit cell and
$\Gamma^\ast = [-\pi,\pi)^3$ the first Brillouin zone of the lattice.

In this paragraph, we introduce two functions $G$ and $W$ which we shall need throughtout the paper. They will respectively yield the so-called direct and exchange terms of periodic Hartree-Fock theory.
We start by introducing the $\Z^3$-periodic Green kernel of the Poisson
interaction~\cite{LS2}, denoted by $G$ and uniquely defined by
$$
\left\{
\begin{array}{l}
\displaystyle -\Delta G = 4\pi \left( \sum_{k \in \Z^3} \delta_k - 1 \right)
\\
\displaystyle G \  \Z^3\mbox{-periodic},\ \mathop{\mbox{min}}_{\R^3} G = 0
\end{array} \right.
$$
and where the first equation holds in the distributional sense.
The Fourier expansion of $G$ is
\begin{equation}
 G(x) = h + \sum_{k \in 2\pi \Z^3 \setminus \left\{0\right\}}
\frac{4\pi}{|k|^2} e^{i k \cdot x}
\label{def_G}
\end{equation}
with $\displaystyle h = \int_\Gamma G > 0$. The electrostatic potential
associated with a $\Z^3$-periodic density $\rho\in L^1_{\rm loc}(\R^3)\cap L^3_{\rm loc}(\R^3)$
 is the $\Z^3$-periodic function defined as
$$
(\rho \ast G)(x) := \int_\Gamma G(x-y) \, \rho(y) \, dy.
$$
We also set for any $\Z^3$-periodic functions $f$ and $g$
\begin{equation*}
D(f,g)  :=  \int_\Gamma \int_\Gamma G(x-y) \, f(x) \, g(y) dx \,
dy.
\end{equation*}

Next we introduce the following function \cite{CLL}
\begin{equation}
W(\eta,z)=\sum_{k\in\Z^3}\frac{e^{ik\cdot\eta}}{|z+k|},\ \ \eta,z\in\R^3.
\label{def_W}
\end{equation}
The function $e^{i\eta\cdot x}W(\eta,x)$ is $\Gamma$-periodic with respect to $x$, when $\eta$ is fixed. So we can write $W$ as a Fourier series and obtain
\begin{equation}
W(\eta,x)=4\pi e^{-i\eta\cdot x}\sum_{k\in2\pi\Z^3}\frac{e^{ik\cdot x}}{|\eta-k|^2}.
\end{equation}
We have that, fixed $x\in\Gamma$, $W(\eta,x)-4\pi\frac{e^{-i\eta\cdot x}}{|\eta|^2}$
is continuous in 0 with respect to $\eta$
and that
\begin{equation}
\lim_{\eta\rightarrow0}\left(W(\eta,x)-4\pi\frac{e^{-i\eta\cdot x}}{|\eta|^2}\right)e^{i\eta\cdot x}+h=G(x).
\end{equation}

\subsection{The periodic Hartree-Fock functional}
The periodic Hartree-Fock functional was studied in \cite{CLL}.

The main object of interest will be the so-called (periodic)
{density matrix} of the electrons. We define the translation operator $\tau_k$ acting on $L^2_{\rm loc}(\R^3)$ as follows: $\tau_k u (x) = u(x-k)$ and introduce the following variational set of density matrices:
\begin{multline*}
\mathcal{P}_{\rm per}=
\bigg\{ \gamma \in {\cS}\big(L^2(\R^3)\big) \; | \; 0 \leq \gamma \leq 1, \
\forall k \in \Z^3,\ \tau_k \gamma = \gamma \tau_k, \\
\int_{\Gamma^*}
\tr_{L^2_\xi(\Gamma)}((1-\Delta_\xi)^{1/2} \gamma_\xi(1-\Delta_\xi)^{1/2}) \, d\xi < \infty  \; \bigg\}.
\end{multline*}
In the whole paper, we use the notation $(A_\xi)_{\xi \in \Gamma^\ast}$ for the Bloch waves decomposition of a periodic operator $A$, see \cite{RSIV,CLL}:
$$
A = \frac{1}{(2\pi)^3}  \int_{\Gamma^\ast} A_\xi \, d\xi,\qquad A_\xi \in \mathcal{S}(L^2_\xi(\Gamma)),
$$
$$L^2_\xi(\Gamma) = \left\{ u \in L^2_{\rm
  loc}(\R^3)\ |\ \tau_k u = e^{-i k \cdot \xi} u, \; \forall k \in \Z^3 \right\}$$
which corresponds to the decomposition in fibers $L^2(\R^3) =
\int^{\oplus}_{\Gamma^*}d\xi L^2_\xi(\Gamma)$.

For any $\gamma\in\cP_{\rm per}$ (and almost every $\xi\in\Gamma^*$), we denote by $\gamma_\xi(x,y)$ the
integral kernel of $\gamma_\xi\in\gS_1(L^2_\xi(\Gamma))$. The density of $\gamma$ is
then the non-negative $\Z^3$-periodic function of $L^1_{\rm loc}(\R^3)
\cap L^3_{\rm loc}(\R^3)$ defined as
$$\rho_\gamma(x):=\frac{1}{(2\pi)^3}\int_{\Gamma^*} \gamma_\xi(x,x) \, d\xi.$$
Notice that for any $\gamma\in\cP_{\rm per}$
$$\int_\Gamma\rho_\gamma(x)dx=\frac{1}{(2\pi)^3}  \int_{\Gamma^\ast}
\tr_{L^2_\xi(\Gamma)}(\gamma_\xi) \, d\xi,$$
i.e. this gives the number of electrons per unit cell. Later we shall add the constraint that the system is neutral and restrict to states $\gamma\in\cP_{\rm per}$ satisfying
$$\int_\Gamma\rho_\gamma(x)dx=Z$$
where $Z$ is the charge of the only nucleus in each unit cell.

The \emph{periodic Hartree-Fock functional} is defined by
\begin{multline}
\cE(\gamma):=\int\limits_{\Gamma^*}\tr_{L^2_\xi(\Gamma)}\left(-\frac12 \Delta_\xi\gamma_\xi\right)\frac{d\xi}{(2\pi)^3}-Z\int_\Gamma G(x)\rho_{\gamma}(x)\,dx\\
+\frac 12 D(\rho_\gamma,\rho_\gamma)-\frac12 \exc(\gamma,\gamma)\label{def_energy}
\end{multline}
for any $\gamma\in\cP_{\rm per}$. In the above formula, $D(\rho_\gamma,\rho_\gamma)/2$ is called the \emph{direct term}, while
$X(\gamma,\gamma)/2$ is called the \emph{exchange term}. The latter is defined for any $\beta,\gamma\in\cP_{\rm per}$ as
\begin{equation}
\exc(\beta,\gamma)=\frac{1}{(2\pi)^6}\iint\limits_{\Gamma^*\times\Gamma^*}d\xi\, d\xi'\iint\limits_{\Gamma\times\Gamma}dx\,dy\
{\beta(\xi,x,y)}W(\xi-\xi',x-y)\overline{\gamma(\xi',x,y)}
\end{equation}
where $W$ is the function defined in \eqref{def_W}.
We remark that $\overline{W(-\eta,z)}=W(\eta,z)=\overline{W(\eta,-z)}$,
so $\exc(\beta,\gamma)=\overline{\exc(\gamma,\beta)}=\exc(\gamma,\beta)$.

In the whole paper, we use the convention that for a density matrix $\gamma\in\cP_{\rm per}$
$$\int\limits_{\Gamma^*}\tr_{L^2_\xi(\Gamma)}\left(-\frac12 \Delta\gamma_\xi\right)\frac{d\xi}{(2\pi)^3}:=\frac12\int\limits_{\Gamma^*}\tr_{L^2_\xi(\Gamma)}\left(\sqrt{-\Delta_\xi} \gamma_\xi\sqrt{-\Delta_\xi}\right)\frac{d\xi}{(2\pi)^3}$$
which is well-defined by assumption.

In \eqref{def_energy}, we have considered that one pointwise nucleus is located at each site on $\Z^3$. We can also consider extended nuclei, in which case the last term of the first line of \eqref{def_energy} is replaced by
$$-\int_\Gamma V_{\rm per}(x)\rho_\gamma(x)\,dx$$
where
$$V_{\rm per}=Z \left(\sum_{k\in\Z^3}\tau_km\right)\ast G,$$
$m$ being a $C^\infty_0(\R^3)$ nonnegative function such that $\int_{\R^3}m(x)=1$ and with a support small enough such that all the $\tau_km$ have disjoint supports. For the sake of simplicity, we shall restrict ourselves to pointwise nuclei, the extension to the smeared nuclei case being straightforward. Of course the pointwise case corresponds to taking $m=\delta_0$.

\subsection{Existence and properties of minimizers}

The following was proved in \cite{CLL}, see Theorem 2.3 p. 698:
\begin{thm}[Existence of minimizers \cite{CLL}]\label{thm_exist} Assume that $Z>0$.
The functional $\cE$ is well-defined and bounded from below on $\cP_{\rm per}$.
Additionally, there exists a minimizer $\gamma\in\cP_{\rm per}$ of the minimization problem
\begin{equation}
I=\inf_{\substack{\gamma\in\cP_{\rm per}\\ \int_\Gamma\rho_\gamma=Z}}\cE(\gamma).
\label{def_min_pb}
\end{equation}
\end{thm}
Notice in \eqref{def_min_pb} we could consider other constraints of the form $\int_\Gamma\rho_\gamma=\lambda$ but this does not make too much physical sense as the periodic system should be neutral in the thermodynamic limit.

\medskip

The purpose of this paper is to show that any minimizer $\gamma$ of \eqref{def_min_pb} solves a specific nonlinear equation. Our main result is the following
\begin{thm}[Self-consistent equation and the last shell]\label{thm_scf} Assume that $Z>0$ and let $\gamma$ be a minimizer of \eqref{def_min_pb}. Then $\gamma$ solves the following nonlinear equation:
\begin{equation}
\left\{\begin{array}{l}
\dps\gamma=\chi_{(-\ii,\mu)}(H_\gamma)+\epsilon\chi_{\{\mu\}}(H_\gamma),\\
\dps(H_\gamma)_\xi=-\Delta_\xi-Z G +\rho_\gamma\ast G-(2\pi)^{-3}\int_{\Gamma^*} W(\xi'-\xi,x-y)\gamma_{\xi'}(x,y)\,d\xi',
\end{array}\right.
\label{eq_SCF}
\end{equation}
where $\epsilon\in\{0,1\}$ and $\mu\in\R$ is a Lagrange multiplier due to the charge constraint $\int_\Gamma\rho_\gamma=Z$.
\end{thm}

\begin{remark}
In the periodic case, for the minimizer $\gamma$ to be a projector, $Z$ need not be an integer as in the atomic setting.
\end{remark}

A result similar to Theorem \ref{thm_scf} was proved for the \emph{reduced model} in \cite[Theorem 1]{CDL}. The reduced model consists in neglecting the exchange term, i.e. the last term of \eqref{def_energy}. In this case, the Hartree-Fock energy becomes convex and it admits a unique minimizer. The proof that it satisfies an equation similar to \eqref{eq_SCF} is then much easier because the only term which depends on $\gamma$ in the operator $H_\gamma$ is $\rho_\gamma\ast G$ when the exchange energy is neglected. Using that $\rho_\gamma\ast G$ is just a periodic function (hence its Bloch decomposition does not depend on $\xi$), one obtains from a result of Thomas \cite{Thomas} that the spectrum of $H_{\gamma}$ is purely absolutely continuous. In particular there cannot be any eigenvalue, hence we can take $\epsilon=0$ in \eqref{eq_SCF}.

When the exchange term is kept in the model as in the present paper, the situation is much more complicated. The main difficulty is that the Bloch decomposition of the last term of $H_\gamma$ depends in a non trivial way of $\xi$. In particular, we do not know if the spectrum of $H_\gamma$ is purely absolutely continuous. It is in principle possible that $\mu$ is an eigenvalue (of infinite multiplicity) of $H_\gamma$ in \eqref{eq_SCF}. However, we are able to prove that any minimizer $\gamma$ is automatically a projector. Theorem \ref{thm_scf} even states that either the minimizer $\gamma$ does not contain the eigenspace corresponding to the eigenvalue $\mu$ ($\epsilon=0$) or it fills it completely ($\epsilon=1$). The fact that $\gamma$ is a projector is an important property which can be used as a basis for the construction of models for crystals with local defects, as this was done in \cite{CDL}.

The rest of the paper is devoted to the proof of Theorem \ref{thm_scf}.

\section{Proof of Theorem \ref{thm_scf}}\label{sec:thm_scf}

\subsubsection*{Step 1. \textit{Properties of $W$.}} We start with the following useful
\begin{lemma}[Properties of $W$] We recall that $W$ is defined in \eqref{def_W}.
We have for all $(\xi,x)\in\Gamma^*\times\Gamma$
\begin{equation}
W(\xi,x)=4\pi\frac{e^{-i\xi\cdot x}}{|\xi|^2}+ e^{-i\xi\cdot x}(G(x)-h)+e^{-i\xi\cdot x}f(\xi,x)
\label{decomp_W_1box}
\end{equation}
where $f$ satisfies
\begin{equation}
\forall\xi,\xi'\in\Gamma^*,\qquad \norm{f(\xi,\cdot)-f(\xi',\cdot)}_{L^\ii(\Gamma)}\leq C|\xi-\xi'|,
\label{f_Lip}
\end{equation}
\begin{equation}
\forall x\in\Gamma,\qquad f(0,x)=0.
\end{equation}
Similarly, we have for all $(\xi,\xi',x,y)\in (\Gamma^*)^2\times\Gamma^2$
\begin{multline}
W(\xi-\xi',x-y)=4\pi \sum_{m\in\Z^3,\ |m|_\ii\leq1} \frac{e^{-i(\xi-\xi'-2\pi m)\cdot(x-y)}}{|\xi-\xi'-2\pi m|^2}\\+ e^{-i(\xi-\xi')\cdot (x-y)}G(x-y)+g(\xi-\xi',x-y)
\label{decomp_W_2boxes}
\end{multline}
where $g\in L^\ii\left((\Gamma^*)^2\times\Gamma^2\right)$ satisfies for all $\xi_1,\xi_1',\xi_2,\xi_2'\in\Gamma^*$
\begin{equation}
 \sup_{x,y\in\Gamma^2}{|g(\xi_1-\xi_1',x-y)-g(\xi_2-\xi_2',x-y)|}\leq C|\xi_1-\xi_1'-\xi_2+\xi_2'|.
\label{regularity_g}
\end{equation}
In \eqref{decomp_W_2boxes}, we have used the notation $|x|_\ii=\sup\{|x_i|,\ i=1,2,3\}$.
\end{lemma}
\begin{proof}
 The proof is contained in \cite[p. 744]{CLL} and we only sketch it for the convenience of the reader. A simple calculation gives
$$f(\xi,x)=4\pi\sum_{m\in\Z^3\setminus\{0\}}e^{2im\cdot x}\frac{4\pi \xi\cdot m-|\xi|^2}{|\xi-2\pi m|^2|2\pi m|^2}.$$
Hence we obtain for $\epsilon$ small enough
$$\norm{f(\xi,\cdot)-f(\xi',\cdot)}_{L^2((1+\epsilon)\Gamma)}^2\leq C|\xi-\xi'|^2.$$
Next one uses that $x\mapsto f(\xi,x)-f(\xi',x)$ is harmonic to infer
$$\norm{f(\xi,\cdot)-f(\xi',\cdot)}_{L^\ii(\Gamma)}\leq C\norm{f(\xi,\cdot)-f(\xi',\cdot)}_{L^1((1+\epsilon)\Gamma)}\leq C|\xi-\xi'|.$$
The proof is the same for $g$.
\end{proof}

In view of the previous result, it is natural to introduce
\begin{align}
 X_G(\gamma,\gamma)&:= \iint_{\Gamma^*\times\Gamma^*}\!\!d\xi\, d\xi'\!\!\iint_{\Gamma\times\Gamma}\!\!dx\, dy\, G(x-y)e^{-i(\xi-\xi' )\cdot(x-y)}\gamma_\xi(x,y)\overline{\gamma_{\xi'}(x,y)}\nonumber\\
&= \iint_{\Gamma\times\Gamma}dx\, dy\, G(x-y)|\tilde\gamma(x,y)|^2
\label{def_exchange_G}
\end{align}
where we have defined $\tilde\gamma_\xi(x,y)=e^{-i\xi\cdot x}\gamma_\xi(x,y)e^{i\xi\cdot y}$. Remark that $\rho_{\gamma}=\rho_{\tilde\gamma}$. Definition \eqref{def_exchange_G}
is quite natural as it is an exchange term taking the same form as in usual Hartree-Fock theory \cite{LS}. For all $\gamma\in\cP_{\rm per}$ we have
\begin{equation}
 |\gamma(x,y)|^2\leq \rho_\gamma(x)\rho_\gamma(y),
\label{estim_echange_density}
\end{equation}
which is a consequence of $\gamma_\xi\geq0$, see, e.g., \cite[p. 746]{CLL}. The same inequality holds with $\gamma$ replaced by $\tilde\gamma$. We infer that
\begin{equation}
 X_G(\gamma,\gamma)\leq D(\rho_\gamma,\rho_\gamma).\label{estim_exchange_G}
\end{equation}
Indeed $\gamma\mapsto D(\rho_\gamma,\rho_\gamma)-X_G(\gamma,\gamma)$ is easily seen to be continuous for the weak topology by Fatou's Lemma, and this can be used to prove the existence of a minimizer.

\subsubsection*{Step 2. \textit{Regularity of the spectral decomposition of $H_\gamma$.}} In the rest of the proof, we consider a minimizer $\gamma$ for \eqref{def_min_pb}. Such a minimizer is known to exist by Theorem \ref{thm_exist}, proved in \cite{CLL}. We know that it has a finite kinetic energy since it belongs to $\cP_{\rm per}$:
$$ \int\limits_{\Gamma^*}\tr_{L^2_\xi(\Gamma)}\left(-\frac12 \Delta_\xi\gamma_\xi\right)\frac{d\xi}{(2\pi)^3}<\ii.$$
We recall the useful inequality \cite[Eq. (4.42)]{CLL}
\begin{equation}
 \int\limits_{\Gamma^*}\tr_{L^2_\xi(\Gamma)}\left(- \Delta_\xi\gamma_\xi\right)\frac{d\xi}{(2\pi)^3}\geq \int_{\Gamma}\left|\nabla\sqrt{\rho_\gamma(x)}\right|^2dx
\label{estim_kinetic_density}
\end{equation}
which proves that the map $\gamma\in\cP_{\rm per}\mapsto\rho_\gamma\in H^1_{\rm per}(\Gamma)\subset L^1_{\rm per}(\Gamma)\cap L^3_{\rm per}(\Gamma)$ is continuous\footnote{This property can be used to properly define the term $D(\rho_\gamma,\rho_\gamma)$ and see that it is continuous for the topology of $\cK$.}.

We investigate the regularity of the eigenvalues and eigenvectors of $H_\gamma$. For simplicity, we study $H_\gamma$ on the fixed Hilbert space $L^2_{\rm per}(\Gamma)$. This means we introduce the unitary operator defined in each Bloch sector by
$$\begin{array}{llll}
U_\xi:& L^2_{\rm per}(\Gamma) & \to& L^2_\xi(\Gamma)\\
 &u(x) & \mapsto & e^{i\xi\cdot x}u(x)
  \end{array}$$
We shall use the convention that when $T_\xi$ is an operator on $L^2_\xi(\Gamma)$, then $\tilde{T}_\xi:=U_\xi^*T_\xi U_\xi$. Let us  denote by $X$ the exchange term defined by its kernel
$$X_\xi(x,y)=(2\pi)^{-3}\int_{\Gamma^*} W(\xi'-\xi,x-y)\gamma_{\xi'}(x,y)\,d\xi'.$$
Then we get using \eqref{decomp_W_2boxes}
\begin{multline}
\tilde X_\xi(x,y)=U_\xi^*X_\xi U_\xi(x,y)=\int_{\Gamma^*}\Bigg[4\pi \sum_{m\in\Z^3,\ |m|_\ii\leq1} \frac{e^{2i\pi m\cdot(x-y)}}{|\xi'-\xi-2\pi m|^2}\\+ G(x-y)+\tilde g(\xi-\xi',x-y)\Bigg]\tilde\gamma_{\xi'}(x,y)\,\frac{d\xi'}{(2\pi)^{3}}
\end{multline}
where we have introduced $\tilde g(\eta,x)=e^{i\eta\cdot x}g(\eta,x)$.
\begin{lemma}\label{lem:exchange_continuous}
The family $(\tilde X_\xi)_{\xi\in\Gamma}$ is bounded in $\cB(L^2_{\rm per}(\Gamma))$ and it is Hölder:
$$\forall 0\leq p<1,\qquad \norm{\tilde X_{\xi_1}-\tilde X_{\xi_2}}_{\cB(L^2_{\rm per}(\Gamma))}\leq C_p|\xi_1-\xi_2|^p.$$
\end{lemma}
\begin{proof}
We have for all $u,v\in L^2_{\rm per}(\Gamma)$
\begin{align*}
&\left|\pscal{(\tilde X_{\xi_1}-\tilde X_{\xi_2})u,v}_{L^2_{\rm per}(\Gamma)}\right|\\
&\quad\leq 4\pi \sum_{m\in\Z^3,\ |m|_\ii\leq1}\int_{\Gamma^*}\left| \frac{1}{|\xi'-\xi_1-2\pi m|^2}-\frac{1}{|\xi'-\xi_2-2\pi m|^2}\right|\times\\
&\qquad\qquad\qquad\qquad\qquad\qquad\qquad\qquad\qquad\times|\pscal{\tilde\gamma_{\xi'}U_{2\pi m}u,U_{2\pi m}v}|\,\frac{d\xi'}{(2\pi)^{3}}\\
&\quad\qquad+\iint_{\Gamma\times\Gamma}dx\,dy\int_{\Gamma^*}\left|\tilde g(\xi'-\xi_1,x-y)-\tilde g(\xi'-\xi_2,x-y)\right|\times\\
&\qquad\qquad\qquad\qquad\qquad\qquad\qquad\qquad\qquad\times|\tilde\gamma_{\xi'}(x,y)|\; |u(y)|\;|v(x)|\,\frac{d\xi'}{(2\pi)^{3}}.
\end{align*}
We recall that $0\leq\gamma\leq1$ which means that $0\leq \tilde\gamma_\xi\leq 1$ for all $\xi\in\Gamma^*$.
Using \eqref{regularity_g} we obtain
\begin{multline*}
\norm{\tilde X_{\xi_1}-\tilde X_{\xi_2}}_{\cB(L^2_{\rm per}(\Gamma))}\leq C|\xi_1-\xi_2|\int_{\Gamma^*}\iint_{\Gamma\times\Gamma}dx\,dy\big|\tilde\gamma_{\xi'}(x,y)\big|^2\,\frac{d\xi'}{(2\pi)^{3}}\\
+ 4\pi \sum_{m\in\Z^3,\ |m|_\ii\leq1}\int_{\Gamma^*}\left| \frac{1}{|\xi'-\xi_1-2\pi m|^2}-\frac{1}{|\xi'-\xi_2-2\pi m|^2}\right|\frac{d\xi'}{(2\pi)^{3}}.
\end{multline*}
As $0\leq \tilde\gamma_\xi\leq 1$ for all $\xi\in\Gamma$, we have
\begin{align*}
(2\pi)^{-3}\int_{\Gamma^*}d\xi'\iint_{\Gamma\times\Gamma}dx\,dy\,\big|\tilde\gamma_{\xi'}(x,y)\big|^2&=\int_{\Gamma^*}\tr_{L^2_{\rm per}(\Gamma)}\big[(\tilde\gamma_{\xi'})^2\big]\,\frac{d\xi'}{(2\pi)^{3}}\\
&\leq \int_{\Gamma^*}\tr_{L^2_{\rm per}(\Gamma)}\big[\tilde\gamma_{\xi'}\big]\,\frac{d\xi'}{(2\pi)^{3}}=Z.
\end{align*}
This easily proves the Hölder regularity.

The proof that $(\tilde X_\xi)_{\xi\in\Gamma}$ is bounded in $\cB(L^2_{\rm per}(\Gamma))$ is essentially the same for all the terms which we have treated above. We just estimate the term involving $G(x-y)$. We notice that
\begin{align*}
&\norm{\int_{\Gamma^*}G(x-y)\tilde\gamma_{\xi'}(x,y)\,\frac{d\xi'}{(2\pi)^{3}}}^2_{\gS_2(L^2_{\rm per}(\Gamma))}\\
&\qquad=\iint_{\Gamma\times\Gamma}dx\,dy\,G(x-y)^2\left|\int_{\Gamma^*}\tilde\gamma_{\xi'}(x,y)\frac{d\xi'}{(2\pi)^3}\right|^2\\
&\qquad= \iint_{\Gamma\times\Gamma}G(x-y)^2\left|\tilde\gamma(x,y)\right|^2dx\,dy\leq \iint_{\Gamma\times\Gamma}G(x-y)^2\rho_\gamma(x)\rho_\gamma(y)dx\,dy
\end{align*}
by \eqref{estim_echange_density}.
The last term is well defined (and it is independent of $\xi$). To see that, we notice that for all $x,y\in\Gamma$
$$G(x-y)=\sum_{m\in\Z^3,\ |m|_\ii\leq1}\frac{1}{|x-y+m|}+h(x-y)$$
where $h(x-y)$ is bounded on $\Gamma\times\Gamma$. We treat for instance the term with $m=0$, the argument being the same for the others. Fix some non negative cut-off function $\chi$ which equals 1 on $\Gamma$ and vanishes outside $(1+\epsilon)\Gamma$ for $\epsilon$ small enough. We have by Hardy's inequality $|x|^{-2}\leq 4(-\Delta)$ and using the periodicity of $\rho_\gamma$ together with the formula $\int_\Gamma\rho_\gamma=Z$
\begin{multline*}
\iint_{\Gamma\times\Gamma}\frac{\rho_\gamma(x)\rho_\gamma(y)}{|x-y|^2}dx\,dy\leq \int_{\R^3}dx\int_{\Gamma}dy\frac{\chi(x)^2\rho_\gamma(x)\rho_\gamma(y)}{|x-y|^2}\\
  \leq 4Z\int_{\R^3}dx\left|\nabla(\chi\sqrt{\rho_\gamma})(x)\right|^2 \leq C\left(Z+\int_{\Gamma}dx\left|\nabla\sqrt{\rho_\gamma}(x)\right|^2\right).
\end{multline*}
Arguing similarly for the other terms and using \eqref{estim_kinetic_density} we get
\begin{multline*}
\norm{\int_{\Gamma^*}G(x-y)\tilde\gamma_{\xi'}(x,y)\,\frac{d\xi'}{(2\pi)^{3}}}^2_{\gS_2(L^2_{\rm per}(\Gamma))}\\
\leq C\left(Z+\int_{\Gamma^*}d\xi'\tr_{L^2_{\xi'}(\Gamma)}\big[(-\Delta)_{\xi'}\gamma_{\xi'}\big]\right)
\end{multline*}
which is independent of $\xi$.
\end{proof}

Recall that the operator $(\tilde{H}_\gamma)_\xi=U_\xi^* (H_\gamma)_\xi U_\xi$ acting on $L^2_{\rm per}(\Gamma)$ reads
\begin{equation}
 (\tilde{H}_\gamma)_\xi=-\frac{\Delta}2-i\xi\cdot\nabla +\frac{|\xi|^2}2-Z G+\rho_\gamma\ast G-\tilde{X}_\xi
\end{equation}
where of course $-\Delta$ and $\nabla$ are respectively the periodic Laplacian and the periodic gradient acting on $L^2_{\rm per}(\Gamma)$.
We denote by $\lambda_k(\xi)$ the eigenvalues of $(\tilde H_\gamma)_\xi$ which are the same as that of $(H_\gamma)_\xi$ since $U_\xi$ is unitary. We may assume that the $\lambda_k(\xi)$ are in nondecreasing order: $\lambda_1(\xi)\leq\lambda_2(\xi)\leq\cdots$.
\begin{lemma}\label{lem:limit} The eigenvalues $\lambda_k(\xi)$ tend to $+\ii$ when $k\to+\ii$, uniformly in $\xi\in\Gamma^*$.
\end{lemma}
\begin{proof}
We have proved in Lemma \ref{lem:exchange_continuous} that $\tilde X_\xi$ is a bounded family of operators in $\cB(L^2_{\rm per}(\Gamma))$. Also we have $\rho_\gamma\geq0$ and $G\geq0$. Hence the following holds on $L^2_{\rm per}(\Gamma)$
$$(\tilde H_\gamma)_\xi\geq -\Delta/2-i\xi\cdot\nabla+|\xi|^2/2 -ZG-C.$$
As $G\in L^2(\Gamma)$, it suffices to apply \cite[Lemma A-2]{Thomas}.
\end{proof}

\begin{lemma}\label{prop_continuity_resolvent} Consider $\Omega$ an open set of $\Gamma^*$ and let $K$ be a compact set in $\C$ such that $\inf_{\xi\in\Omega}{\rm d}(K,\sigma(\tilde{H}_\gamma)_\xi)\geq \epsilon>0$. Then we have that
\begin{enumerate}
\item $(1-\Delta)\left((\tilde{H}_\gamma)_\xi-z\right)^{-1}$ is bounded on $L^2_{\rm per}(\Gamma)$, uniformly with respect to $\xi\in\Omega$ and $z\in K$;
\item the map $\xi\mapsto(1-\Delta)\left((\tilde{H}_\gamma)_\xi-z\right)^{-1}\in \cB(L^2_{\rm per}(\Gamma))$ is Hölder with respect to $\xi\in\Omega$, uniformly in $z\in K$.
\end{enumerate}
\end{lemma}
\begin{proof}
By Lemma \ref{lem:limit}, we know that there exists a real number $z_0>0$ such that $\text{d}\big(-z_0,\sigma(\tilde H_\gamma)_\xi\big)>\epsilon$ for all $\xi\in \Gamma^*$ and some $\epsilon>0$. We choose a constant $c>z_0$ large enough. We have
$$\left((\tilde{H}_\gamma)_\xi+c\right)(c-\Delta)^{-1}=1/2+R_c,$$
where
$$R_c:=-i\xi\cdot\frac{\nabla}{c-\Delta}+|\xi|^2/2(c-\Delta)^{-1}+\left(\rho_\gamma\ast G- ZG-X_\xi\right)(c-\Delta)^{-1}.$$
We have
$$\norm{\xi\cdot\frac{\nabla}{c-\Delta}}_{\cB(L^2_{\rm per}(\Gamma))}+\norm{|\xi|^2/2(c-\Delta)^{-1}}_{\cB(L^2_{\rm per}(\Gamma))} \leq \frac{C}{1+c}$$
and, as $X_\xi$ is uniformly bounded in $\cB(L^2_{\rm per}(\Gamma))$ by Lemma \ref{lem:exchange_continuous},
$$\norm{X_\xi(c-\Delta)^{-1}}_{\cB(L^2_{\rm per}(\Gamma))} \leq \frac{C}{1+c}.$$
Next we can use that $G\in L^2_{\rm per}(\Gamma)$ to get that
$$\norm{G(x)(c-\Delta)^{-1}}^2_{\gS_2(L^2_{\rm per}(\Gamma))}= \norm{G}^2_{L^2_{\rm per}(\Gamma)}\norm{(c+|\cdot|^2)^{-1}}^2_{\ell^2(2\pi\Z^3)}\leq \frac{C}{c^{1/4}}.$$
Similarly we have $G\in L^1_{\rm per}(\Gamma)$ and $\rho_\gamma\in L^2_{\rm per}(\Gamma)$, hence $\rho_\gamma\ast G\in L^2_{\rm per}(\Gamma)$ and the same estimate furnishes the bound
$$\norm{(\rho_\gamma\ast G)(x)(c-\Delta)^{-1}}^2_{\gS_2(L^2_{\rm per}(\Gamma))}\leq \frac{C}{c^{1/4}}.$$
Eventually we have proved that
$$\norm{R_c}_{\cB(L^2_{\rm per}(\Gamma))}\leq \frac{C}{c^{1/4}}.$$
This shows that for $c$ large enough, the operator $((\tilde{H}_\gamma)_\xi+c)(c-\Delta)^{-1}$ is invertible and that its inverse is bounded:
\begin{equation}
(c-\Delta)\left((\tilde{H}_\gamma)_\xi+c\right)^{-1}\in\cB(L^2_{\rm per}(\Gamma)).
\label{resolvent_bounded}
\end{equation}
The last step is to use the formula
$$\left((\tilde{H}_\gamma)_\xi-z\right)^{-1}=\left((\tilde{H}_\gamma)_\xi+c\right)^{-1}+(z+c)\left((\tilde{H}_\gamma)_\xi+c\right)^{-1}\left((\tilde{H}_\gamma)_\xi-z\right)^{-1}$$
which, when inserted in \eqref{resolvent_bounded} yields that $(1-\Delta)((\tilde{H}_\gamma)_\xi-z)^{-1}$ is bounded on $L^2_{\rm per}(\Gamma)$, uniformly with respect to $\xi\in\Omega$ and $z\in K$.

For the second point of the Lemma, we use the resolvent formula
\begin{align*}
&(1-\Delta)\left[\left((\tilde{H}_\gamma)_\xi-z\right)^{-1}-\left((\tilde{H}_\gamma)_{\xi'}-z\right)^{-1}\right]=\\
&\qquad=(1-\Delta)\left((\tilde{H}_\gamma)_\xi-z\right)^{-1}\left((\tilde{H}_\gamma)_\xi- (\tilde{H}_\gamma)_{\xi'}\right)\left((\tilde{H}_\gamma)_{\xi'}-z\right)^{-1}=\\
&\qquad=(1-\Delta)\left((\tilde{H}_\gamma)_\xi-z\right)^{-1}\left(-i(\xi-\xi')\cdot\nabla+|\xi|^2-|\xi'|^2-\tilde{X}_\xi+\tilde{X}_{\xi'}\right)\times\\
&\qquad\qquad\qquad\qquad\qquad\qquad \times(1-\Delta)^{-1}(1-\Delta)\left((\tilde{H}_\gamma)_{\xi'}-z\right)^{-1}.
\end{align*}
Next we can use that $\xi\mapsto \tilde X_\xi$ is Hölder for the norm of $\cB(L^2_{\rm per}(\Gamma))$ as shown in Lemma \ref{lem:exchange_continuous}, and that $\nabla(1-\Delta)^{-1}$ is a bounded operator on $L^2_{\rm per}(\Gamma)$.
\end{proof}

\begin{lemma}\label{lem:continuity}
The eigenvalues $\lambda_k(\xi)$ are Hölder with respect to $\xi\in\Gamma$.
\end{lemma}
\begin{proof}
As before we take $c$ large enough such that $\text{d}\big(-c,\sigma(\tilde H_\gamma)_\xi\big)>\epsilon$ for all $\xi\in \Gamma^*$ and some $\epsilon>0$. For every $\xi\in\Gamma^*$, the spectrum of the self-adjoint operator $R(\xi):=((\tilde{H}_\gamma)_\xi+c)^{-1}$ is composed of the eigenvalues $\mu_k(\xi):=(\lambda_k(\xi)+c)^{-1}$. By Lemma \ref{prop_continuity_resolvent}, $\xi\mapsto R(\xi)$ is a Hölder family in $\cB(L^2_{\rm per}(\Gamma))$. By the usual min-max Courant-Fisher formula, we have
$$|\mu_k(\xi)-\mu_k(\xi')|\leq \norm{R(\xi)-R(\xi')}$$
hence for any $k\geq1$, $\xi\in\Gamma^*\mapsto\mu_k(\xi)$ is a Hölder function. This \emph{a fortiori} proves the same property for the eigenvalues $\lambda_k(\cdot)$.
\end{proof}

\begin{lemma}\label{cor_eigenvectors}
Let $\Omega$ be an open subset of $\Gamma^*$ and $I=(a,b)$ an interval of $\R$ such that $\sigma(\tilde{H}_\gamma)_{\xi}\cap\{a,b\}=\emptyset$ for all $\xi\in\Omega$. Then the map
$$\xi\in\Omega\mapsto (1-\Delta)\chi_{I}(\tilde{H}_\gamma)_{\xi}\in\cB(L^2_{\rm per}(\Gamma))$$
is Hölder. In particular, we can find an orthonormal basis $(u_1(\xi),...,u_K(\xi))$ of the range of $\chi_{I}(\tilde{H}_\gamma)_{\xi}$ such that $\xi\in\Omega \mapsto u_k(\xi)\in H^2_{\rm per}(\Gamma)$ is Hölder with respect to $\xi\in \Omega$.
\end{lemma}
\begin{proof}
 This is a simple application of Cauchy's formula \cite{Kato}
\begin{equation}
 \chi_{I}(\tilde{H}_\gamma)_{\xi}=\frac1{2i\pi}\oint_\cC dz \left((\tilde{H}_\gamma)_\xi-z\right)^{-1}
\label{Cauchy}
\end{equation}
where $\cC$ is a smooth curve in $\C$ enclosing the interval $I$ and intersecting the real axis at $a$ and $b$ only.
\end{proof}

\subsubsection*{Step 3. \textit{Variational property of a minimizer $\gamma$.}}
We follow a well-known argument \cite{Bach,BLS} (see \cite[Section 4]{LSY} for a very similar setting) and consider a fixed state $\gamma'\in\cP_{\rm per}^Z:=\{\gamma\in\cP_{\rm per}\ |\ \int_\Gamma\rho_\gamma=Z\}$. As $\cP_{\rm per}^Z$ is convex, we have $(1-t)\gamma+t\gamma'\in \cP_{\rm per}^Z$ for all $t\in[0,1]$. Hence, since $\gamma$ is a minimizer
\begin{equation}
\forall t\in[0,1],\qquad \frac{E(\gamma+t(\gamma'-\gamma))-E(\gamma)}{t}\geq0.
\end{equation}
Expanding, we obtain
\begin{multline}
\int\limits_{\Gamma^*}\tr_{L^2_\xi(\Gamma)}\left((H_\gamma)_\xi(\gamma'-\gamma)_\xi\right)\frac{ d\xi}{(2\pi)^3}\ +\\+\frac{t}{2}\bigg\{D(\rho_{\gamma'-\gamma},\rho_{\gamma'-\gamma})-X(\gamma'-\gamma,\gamma'-\gamma)\bigg\}\geq0
\label{2nd_order_condition}
\end{multline}
where $H_\gamma$ is the mean-field operator defined in \eqref{eq_SCF}.
Taking $t=0$, we obtain that $\gamma$ is also a minimizer of the linearized functional at $\gamma$,
$$\gamma'\mapsto \int\limits_{\Gamma^*}\tr_{L^2_\xi(\Gamma)}\left((H_\gamma)_\xi\gamma'_\xi\right)\frac{ d\xi}{(2\pi)^3},$$
on the convex set $\cP_{\rm per}^Z$. For any $\xi\in \Gamma^*$ we can choose $\{\varphi_k(\xi,\cdot)\}_k$
a basis of eigenfunctions of $\left(H_{\gamma}\right)_\xi$ with eigenvalues $\lambda_k(\xi)$, such that
\begin{equation}
(H_\gamma)_\xi=\sum_{k\geq1}\lambda_k(\xi)\ket{\phi_k(\xi)}\bra{\phi_k(\xi)}.
\end{equation}
We know from Lemma \ref{lem:continuity} that each $\lambda_k(\xi)$ is Hölder with respect to $\xi$.
Let us introduce like in \cite[Appendix]{CDL} the function
$$C:\kappa\mapsto \sum_{k\geq1} \left|\{\xi\in\Gamma^*\ |\ \lambda_k(\xi)\leq\kappa\}\right|.$$
The function $C$ is nondecreasing on $\R$. The operator $H_\gamma$ being bounded from below, we have $C\equiv0$ on $(-\ii,\inf\lambda_k(\Gamma^*))$. Also we know that $\lim_{\kappa\to\ii}C(\kappa)=+\ii$ by Lemma \ref{lem:limit}.
We deduce that there exists a $\mu\in\R$ and a periodic operator $\delta\in\cP_{\rm per}$ such that
$$\lim_{\kappa\to\mu^-}C(\kappa)\leq Z\leq \lim_{\kappa\to\mu^+}C(\kappa)$$
and
$$\gamma=\chi_{(-\ii,\mu)}(H_\gamma)+\delta$$
where $0\leq\delta\leq 1$ and $\text{Ran}(\delta)\subset\ker(H_\gamma-\mu)$.
The proof of this fact was given in the Appendix of \cite{CDL}.
When $\mu$ is not an eigenvalue of $H_\gamma$, i.e. $\left|\{\xi\in\Gamma^*\ |\ \exists k,\ \lambda_k(\xi)=\mu\}\right|=0$, the proof of Theorem \ref{thm_scf} is finished, as necessarily $\delta=0$.
What rests to prove is that if $\mu$ is an eigenvalue of $H_\gamma$, then $\delta=0$ or $\delta=\chi_{\{\mu\}}(H_\gamma)$.

\subsubsection*{Step 4. \textit{The Fermi level is either empty or totally filled.}}
Now we argue by contradiction and assume that $\mu$ is an eigenvalue of $H_\gamma$:
$$|\{\xi\in\Gamma^*\ |\ \exists k\geq1,\ \lambda_k(\xi)=\mu\}|\neq0.$$
We also assume that $\delta\neq0$ and $\delta\neq \chi_{\{\mu\}}(H_\gamma)$. The following lemma will be a key result to construct perturbations of $\delta$.

\begin{lemma}\label{lem:perturbation}Assume that $\mu$ is an eigenvalue of $H_\gamma$, that $\delta\neq 0$ and $\delta\neq\chi_{\{\mu\}}(H_\gamma)$.
Then there exists a constant $\epsilon>0$, a Borel set $A\subseteq\Gamma^*$ with $|A|\neq0$ and two continuous functions $\xi\in A\mapsto u(\xi)\in H^2_{\rm per}(\Gamma)$ and $\xi\in A\mapsto u'(\xi)\in H^2_{\rm per}(\Gamma)$ such that
$$ \forall \xi\in A,\qquad u(\xi),u'(\xi)\in \ker((\tilde H_\gamma)_\xi-\mu),$$
$$\norm{u(\xi)}_{L^2_{\rm per}(\Gamma)}=\norm{u'(\xi)}_{L^2_{\rm per}(\Gamma)}=1$$
and, denoting $\phi(\xi)=U^*_\xi u(\xi)$ and $\phi'(\xi)=U^*_\xi u'(\xi)$,
\begin{equation}
0\leq \delta_\xi+t\ket{\phi(\xi)}\bra{\phi(\xi)}-t'\ket{\phi'(\xi)}\bra{\phi'(\xi)}\leq 1
\label{perturbation}
\end{equation}
on $L^2_\xi(\Gamma)$, for all $\xi\in A$ and all $t,t'\in[0,\epsilon)$.
\end{lemma}
\begin{proof}
 Assuming the eigenvalues are in nondecreasing order and using that $\lambda_k(\xi)\to\ii$ as $k\to\ii$, uniformly in $\xi$ and Lemma \ref{lem:continuity}, we deduce that there exists $k_1$, $m$, $\epsilon'>0$ and a subset $A$ of $\Gamma^*$ with $|A|\neq0$ such that
\begin{equation}
\lambda_{k_1-1}(\xi) \leq \mu-\epsilon'<\lambda_{k_1}(\xi)=\mu=\lambda_{k_1+m-1}(\xi)<\mu+\epsilon'\leq \lambda_{k_n+m}(\xi),
\label{condition_vp}
\end{equation}
\begin{equation}
0<\epsilon'\leq \tr_{L^2_\xi(\Gamma)}(\Pi_\xi\gamma_\xi\Pi_\xi)\leq m-\epsilon'
\label{condition_trace}
\end{equation}
for all $\xi\in A$ and where we have introduced $\Pi_\xi$, the orthogonal projector on $\ker((H_\gamma)_\xi-\mu)$ in $L^2_\xi(\Gamma)$. As $0\leq\gamma_\xi\leq1$ for all $\xi\in\Gamma^*$ and $\Pi_\xi\gamma_\xi\Pi_\xi=\delta_\xi$, \eqref{condition_trace} means exactly that $\delta_\xi$ is neither 0 nor 1 when $\xi\in A$.

Now we choose an adequate basis of the range of $\Pi_\xi$. By Lemma \ref{lem:continuity}, we know that the eigenvalues $\lambda_k(\xi)$ of $(H_\gamma)_\xi$ depend continuously on $\xi$. Hence \eqref{condition_vp} is satisfied on the set $\overline{A}$. As $|A|\neq0$, we deduce that the interior of $\overline{A}$ is not empty. Hence, decreasing $A$ if necessary, we can assume without any loss of generality that $A\subseteq\Omega$ where $\Omega$ is an open set on which \eqref{condition_vp} holds true\footnote{However we cannot assume a similar property for \eqref{condition_trace} because we have no information on the regularity of $\gamma_\xi$ on the range of $\Pi_\xi$.}. By Lemma \ref{cor_eigenvectors}, we can choose  $u_{k_1}(\xi),...,u_{k_1+m-1}(\xi)$ an orthonormal basis in $L^2_{\rm per}(\Gamma)$ of $\ker((\tilde H_\gamma)_\xi-\mu)$ for all $\xi$ in $\Omega$ such that $\xi\mapsto u_k(\xi)$ is a Hölder map in $H^2_{\rm per}(\Gamma)$ on $\Omega$. Let us recall that $(\tilde H_\gamma)_\xi:=U_\xi^*(H_\gamma)_\xi U_\xi$ where $U_\xi:L^2_{\rm per}(\Gamma)\to L^2_\xi(\Gamma)$ is the unitary operator which acts as a multiplication by the function $e^{i\xi\cdot x}$. Of course the functions $\phi_k(\xi):=U^*_\xi u_k(\xi)$ are eigenvectors of $(H_\gamma)_\xi$ with eigenvalue $\mu$, by definition.

Next, for any $\xi\in A$, we may introduce the $m\times m$ matrix $M_\xi$ of $\Pi_\xi\delta_\xi\Pi_\xi$ in the basis $\{\phi_k(\xi)\}_{k=k_1}^{k_1+m-1}$
$$(M_\xi)_{ij}=\pscal{\phi_{k_1-1+i}(\xi),\delta_\xi\phi_{k_1-1+j}(\xi)}_{L^2_\xi(\Gamma)}.$$
As $\xi\mapsto M_\xi\in\cS(\C^m)$ is a bounded measurable function, Lusin's theorem \cite{Rudin} tells us that it is continuous on a Borel set $B\subset \Omega$ with $|\Omega\setminus B|$ as small as we want. Hence, replacing $A$ by $A\cap B$ if necessary, we may assume that $\xi\mapsto M_\xi\in\cS(\C^m)$ is continuous on $A$.
The eigenvalues $0\leq n_1(\xi)\leq\cdots \leq n_m(\xi)\leq 1$ of the matrix $M_\xi$ are then continuous functions on $A$. Our assumption is that on $A$, $M_\xi$ is not the zero matrix and is not the identity matrix.

First we treat the case when $M_\xi$ has an eigenvalue which is not equal to 0 and not equal to 1. This means decreasing $A$ one more time if necessary, we may assume by continuity that for all $\xi\in A$
$$n_{k}(\xi)<\epsilon\leq n_{k+1}(\xi)\leq\cdots\leq n_{k+J_1}(\xi)\leq 1-\epsilon<n_{k+J_1+1}(\xi)$$
with the convention that $n_{0}=-\ii$ and $n_{m+1}=\ii$.
By Cauchy's formula \eqref{Cauchy} the projector on $\bigoplus_{j=1}^{J_1}\ker(M_\xi-n_{k+j})$ depends continuously on $\xi\in A$, see \cite{Kato}. Hence we can find a continuous function $\xi\in A\mapsto v(\xi)\in \bigoplus_{j=1}^{J_1}\ker(M_\xi-n_{k+j})\subset \C^m$ such that
\begin{equation}
\forall t\in(-\epsilon,\epsilon),\qquad 0\leq M_\xi+t v(\xi) v(\xi)^*\leq 1.
\label{first_case}
\end{equation}
and $|v(\xi)|=1$ for all $\xi\in A$. We may then introduce
$$\phi(\xi)=\sum_{j=1}^{m}v(\xi)_j\;\phi_{k_1-1+j}(\xi)\in L^2_\xi(\Gamma)\quad \text{and}\quad u(\xi)=U_\xi\phi(\xi)\in L^2_{\rm per}(\Gamma).$$
Notice that the function $\xi\in A\mapsto u(\xi)\in H^2_{\rm per}(\Gamma)$ is continuous by the definition of $v(\xi)$ and $\phi_k(\xi)$. To get \eqref{perturbation}, we just use $\phi':=\phi$ and $u':=u$.

The other case is when $M_\xi$ is always a projector for almost every $\xi\in A$. By continuity of $\xi\mapsto\tr(M_\xi)$ we may, decreasing $A$ one more time if necessary, assume that it has a constant rank $0<J_2<m$. This means we have on $A$ $$0=n_1(\xi)=\cdots=n_{m-J'}(\xi)<n_{m-J_2+1}(\xi)=\cdots=n_m(\xi)=1.$$
As before by Cauchy's formula \eqref{Cauchy} the projectors on $\ker(M_\xi)$ and on $\ker(M_\xi-1)$ depend continuously on $\xi\in A$. Hence we can find two continuous functions $\xi\in A\mapsto v(\xi)\in \ker(M_\xi)$ and $\xi\in A\mapsto v'(\xi)\in \ker(M_\xi-1)$ such that
$$\forall t,t'\in[0,\epsilon),\quad 0\leq M_\xi+t v(\xi) v(\xi)^*-t' v'(\xi) v'(\xi)^*\leq 1$$
and $|v(\xi)|=|v'(\xi)|=1$ for all $\xi\in A$. It now rests to define
$$\phi(\xi):=\sum_{j=1}^{m}v(\xi)_j\;\phi_{k_1-1+j}(\xi)\in L^2_\xi(\Gamma),$$
$$\phi'(\xi):=\sum_{j=1}^{m}v'(\xi)_j\;\phi_{k_1-1+j}(\xi)\in L^2_\xi(\Gamma).$$
Equation \eqref{perturbation} holds with $u(\xi)=U_\xi\phi(\xi)$ and $u'(\xi)=U_\xi\phi'(\xi)$.
\end{proof}

\begin{lemma}\label{lem:condition_positive}Assume that $\mu$ is an eigenvalue of $H_\gamma$, that $\delta\neq 0$ and $\delta\neq\chi_{\{\mu\}}(H_\gamma)$, and let be $A$ and $\phi(\xi)$, $\phi'(\xi)$ as in Lemma \ref{lem:perturbation}. Then for all $\eta,\eta' \in L^\ii(A,\R^+)$ such that $\int_A\eta=\int_A\eta'$, we have
\begin{equation}
D(\rho_{R},\rho_{R})-X(R,R)\geq0
\label{cond_positive}
\end{equation}
where $R$ is the periodic operator defined by
$$R_\xi=\eta(\xi)\ket{\phi(\xi)}\bra{\phi(\xi)}-\eta'(\xi)\ket{\phi'(\xi)}\bra{\phi'(\xi)}.$$
\end{lemma}
\begin{proof}
Let $\eta,\eta' \in L^\ii(A,\R^+)$ be such that  $\int_A(\eta-\eta')=0$ and $R$ as defined above. First we remark that for $0\leq(\norm{\eta}_{L^\ii}+\norm{\eta'}_{L^\ii})t<\epsilon$, we have  $\gamma+tR\in\cP_{\rm per}^Z$ by construction of $\phi(\xi)$ and $\phi'(\xi)$.
Next we use Equation \eqref{2nd_order_condition} with $\gamma'=\gamma+tR$. We compute
\begin{align*}
\int\limits_{\Gamma^*}\tr_{L^2_\xi(\Gamma)}\left[(H_{\gamma})_\xi\; R_\xi\right] \frac{d\xi}{(2\pi)^3}&=\int\limits_{A}\eta(\xi)\pscal{(H_{\gamma})_\xi\phi(\xi),\phi(\xi)}_{L^2_\xi(\Gamma)} \frac{d\xi}{(2\pi)^3}\\
&\quad-\int\limits_{A}\eta'(\xi)\pscal{(H_{\gamma})_\xi\phi'(\xi),\phi'(\xi)}_{L^2_\xi(\Gamma)} \frac{d\xi}{(2\pi)^3}\\
&=\mu
\int\limits_{A}(\eta(\xi)-\eta'(\xi))\frac{d\xi}{(2\pi)^3}=0,
\end{align*}
hence the first order term of \eqref{2nd_order_condition} vanishes and the result follows.
\end{proof}

The last step is to construct functions $\eta,\eta'\in L^\ii(A,\R^+)$ such that $\int_A(\eta-\eta')=0$ and $D(\rho_{R},\rho_{R})-X(R,R)<0$ with $R$ defined as above. This will contradict \eqref{cond_positive} and finish the proof. Indeed, we even prove that
$$\inf_{\substack{\eta,\eta'\in L^\ii(A,\R^+),\\ \int_A(\eta-\eta')=0}}\big( D(\rho_{R},\rho_{R})-X(R,R)\big)=-\ii.$$
The idea of the proof is somewhat similar to that of the atomic case \cite{Lieb,Bach,BLLS,BLS,LSY}. The role of the perturbation $R$ is to transfer some charge from the eigenvector $\phi(\xi)$ to the eigenvector $\phi'(\xi)$ within the last level $\mu$. The originality of the periodic case studied in the present paper is that the transfer needs to be done between two different Bloch sectors $L^2_{\xi_1}(\Gamma)$ and $L^2_{\xi_2}(\Gamma)$. This means essentially we want to take $\eta=\delta_{\xi_1}$ and $\eta'=\delta_{\xi_2}$ in Lemma \ref{lem:condition_positive}.

As $|A|\neq0$, we may find two points $\xi_1$ and $\xi_2$ in $A$ such that $|A\cap B(\xi_1,\lambda)|\neq0$ and $|A\cap B(\xi_2,\lambda)|\neq0$ for all $\lambda>0$. Here $B(\xi,\lambda)$ denotes the ball of radius $\lambda$ centered at $\xi$. We may also assume that $0<|\xi_1-\xi_2|\leq 1/4$ in such a way that $\xi-\xi'\in\Gamma^*$ if $\xi\in B(\xi_1,\lambda)$, $\xi'\in B(\xi_2,\lambda)$ and $\lambda$ is small enough. Next we define
$$\eta_\lambda=\frac{\1_{A\cap B(\xi_1,\lambda)}}{|A\cap B(\xi_1,\lambda)|},\qquad \eta'_\lambda=\frac{\1_{A\cap B(\xi_2,\lambda)}}{|A\cap B(\xi_2,\lambda)|}$$
in such a way that $\eta_\lambda\wto \delta_{\xi_1}$ and $\eta'_\lambda\wto \delta_{\xi_2}$ weakly as $\lambda\to0$. We denote by $R^\lambda$ the associated periodic operator, the Bloch decomposition of which is given by
$$(R^\lambda)_\xi=\eta_\lambda(\xi)\ket{\phi(\xi)}\bra{\phi(\xi)}-\eta'_\lambda(\xi)\ket{\phi'(\xi)}\bra{\phi'(\xi)}.$$
We also introduce as usual the family of operators acting on the fixed space $L^2_{\rm per}(\Gamma)$
$$(\tilde R^\lambda)_\xi=\eta_\lambda(\xi)\ket{u(\xi)}\bra{u(\xi)}-\eta'_\lambda(\xi)\ket{u'(\xi)}\bra{u'(\xi)}.$$

Next using \eqref{decomp_W_1box}, we get
$$(2\pi)^6\bigg(D(\rho_{R^\lambda},\rho_{R^\lambda})-X(R^\lambda,R^\lambda)\bigg)=I_0(\lambda)+I_1(\lambda)+I_2(\lambda)$$
where
\begin{align*}
I_0(\lambda)&=(2\pi)^6\bigg(D(\rho_{\tilde R^\lambda},\rho_{\tilde R^\lambda})-X_G(\tilde R^\lambda,\tilde R^\lambda)\bigg)\\
&\quad+h\iint\limits_{\Gamma\times\Gamma}dx\, dy\,\left|\int\limits_{A}\left(\eta_\lambda(\xi)u(\xi,x)\overline{u(\xi,y)}-\eta'_\lambda(\xi)u'(\xi,x)\overline{u'(\xi,y)}\right)d\xi\right|^2\\
&\quad-\iint\limits_{\Gamma\times\Gamma}dx\, dy\,\iint\limits_{A\times A}d\xi\, d\xi'\,f(\xi-\xi',x-y)\times\\
&\qquad\qquad\qquad\qquad \times \left(\eta_\lambda(\xi)u(\xi,x)\overline{u(\xi,y)}-\eta'_\lambda(\xi)u'(\xi,x)\overline{u'(\xi,y)}\right)\times\\
&\qquad\qquad\qquad\qquad \times \left(\eta_\lambda(\xi')\overline{u(\xi',x)}{u(\xi',y)}-\eta'_\lambda(\xi')\overline{u'(\xi',x)}{u'(\xi',y)}\right)\\
&\quad+8\pi\iint\limits_{A\times A}d\xi\, d\xi'\,
\frac{\eta_\lambda(\xi)\eta'_\lambda(\xi')}{|\xi-\xi'|^2}\left|\pscal{u(\xi),u'(\xi')}_{L^2_{\rm per}(\Gamma)}\right|^2,
\end{align*}
$$I_1(\lambda) =
-4\pi\iint\limits_{A\times A}d\xi\, d\xi'\,
\frac{\eta_\lambda(\xi)\eta_\lambda(\xi')}{|\xi-\xi'|^2}\left|\pscal{u(\xi),u(\xi')}_{L^2_{\rm per}(\Gamma)}\right|^2,$$
$$I_2(\lambda) =
-4\pi\iint\limits_{A\times A}d\xi\, d\xi'\,
\frac{\eta'_\lambda(\xi)\eta'_\lambda(\xi')}{|\xi-\xi'|^2}\left|\pscal{u'(\xi),u'(\xi')}_{L^2_{\rm per}(\Gamma)}\right|^2.$$
To finish the proof, it rests to show the following
\begin{lemma}
 We have
$$\lim_{\lambda\to0} I_0(\lambda)=I_0\in\R,\qquad\lim_{\lambda\to0} I_1(\lambda)=\lim_{\lambda\to0} I_2(\lambda)=-\ii.$$
\end{lemma}
\begin{proof}
Using the important property that the maps $\xi\in\Gamma^*\mapsto u(\xi)\in H^2_{\rm per}(\Gamma)$ and $\xi\in\Gamma^*\mapsto u'(\xi)\in H^2_{\rm per}(\Gamma)$ are continuous by construction, we see that
\begin{align*}
\lim_{\lambda\to0}I_0(\lambda)&=(2\pi)^6\iint\limits_{\Gamma\times\Gamma}dx\, dy\,G(x-y)\left(Z(x,x)Z(y,y)-|Z(x,y)|^2\right)\\
&+2\Re\iint\limits_{\Gamma\times\Gamma}dx\, dy\,\left(f(\xi_1-\xi_2,x-y)+\frac{4\pi}{|\xi_1-\xi_2|^2}\right)u(\xi_1,x)\overline{u(\xi_1,y)}\times\\
&\qquad \times\overline{u'(\xi_2,x)}u'(\xi_2,y)+h\iint\limits_{\Gamma\times\Gamma}dx\, dy|Z(x,y)|^2
\end{align*}
with
$$Z(x,y):=u(\xi_1,x)\overline{u(\xi_1,y)}-u'(\xi_2,x)\overline{u'(\xi_2,y)}.$$
We have also used that $f(0,x)=0$.
Finally, we use again the continuity of $\xi\mapsto u(\xi)$ in $L^2_{\rm per}(\Gamma)$ to infer that
$$\pscal{u(\xi'),u(\xi)}_{L^2_{\rm per}(\Gamma)}=1+o_{\lambda\to0}(1)$$
when $\xi,\xi'\in B(\xi_1,\lambda)$.
Hence, for $\lambda$ small enough
$$I_1(\lambda)\leq -2\pi\iint\limits_{A\times A}d\xi\, d\xi'\,
\frac{\eta^\lambda_1(\xi)\eta^\lambda_1(\xi')}{|\xi-\xi'|^2}\leq -\frac{2\pi}{\lambda^2}$$
which proves that $\lim_{\lambda\to0}I_1(\lambda)=-\ii$. The argument is the same for $I_2(\lambda)$.
\end{proof}

\noindent{\bf Acknowledgement.} We would like to thank \'Eric
S\'{e}r\'{e} and Isabelle Catto for very useful discussions.

\addcontentsline{toc}{section}{References}
\bibliographystyle{amsplain}

\end{document}